\documentclass[review]{elsarticle}
\usepackage{lineno,hyperref}
\modulolinenumbers[5]

\usepackage{xcolor}
\usepackage{ragged2e}
\usepackage{graphicx}
\usepackage{pbox}
\usepackage{amsmath}
\usepackage{tabularx}
\usepackage{subfig}
\usepackage{graphicx}
\usepackage{amsmath, amsfonts, bm, color, ulem}
\usepackage{units}

\usepackage{comment}
\usepackage[english]{babel}

\makeatletter
\def\ps@pprintTitle{%
  \let\@oddhead\@empty
  \let\@evenhead\@empty
  \def\@oddfoot{\reset@font\hfil\thepage\hfil}
  \let\@evenfoot\@oddfoot
}
\makeatother

\begin{document}

\begin{frontmatter}

\title{Two-dimensional array for operating an oscillator in Euclidean space $\mathbb{R}^3$, as a power multiplier}



\author[1]{Elena Campillo Abarca}
\author[1]{Jimena de Hita Fernández,}
\author[1]{Laura Morón Conde}
\author[1]{Almudena Martínez Cedillo,}
\author[1]{Miguel León Pérez}
\author[1]{Andrei Sipos}
\author[1]{Daniel Heredia Doval}
\author[1]{Javier Domingo Serrano}
\author[1]{Rubén González Martínez}

\address[1]{BioCoRe, Biological Cooperative Research S.Coop, Madrid, Spain}



\begin{abstract}

In the present study an oscillator system formed by a seesaw connected to a simple pendulum coupled to a mobile platform with a certain slope, is analyzed. The observed properties of the system when faced with a possible displacement of the mobile are affected in terms of energy loss, which can be reflected in the angle, the variation of apparent weight and the height. The possible variations which can be introduced modify these parameters, so that the system tries to compensate them in order to preserve its symmetry. The working of this system can be described by means of a transformation matrix where the nature of the movement is represented. Such tool enables us to observe the variations that the system may experience as rotation and translation functions. Therefore, the matrix offers a clear visualization of the values expressed and thus of the need to either provide or extract energy to or from the system.

\end{abstract}

\end{frontmatter}

\section{Introduction}

The simple pendulum system can illustrate a wide variety of physical behaviors \cite{bender2006complex}, which are extended if combined with other assumptions \cite{stachowiak2006numerical, yi2000stabilization}. In previous works \cite{ElenayJimena}, the behavior of an oscillator formed by a pendulum connected to a seesaw has been analyzed. This assumption has allowed us to introduce the notion of apparent weight, showing how it is affected depending on the geometry of the system. In this work we study what happens to the motion of the oscillator when its pendulum is coupled to a moving platform with a certain slope.
Specifically, the research will analyze the forces exerted on the point mass, how it affects its motion and the variation of apparent weight, and how it would be if it were in an ideal situation. The geometrical characteristics of the system will reveal the interdependencies that relate some parameters to others.

Likewise, the symmetry of the system allows recourse to Noether's theorem (1918) \cite{noether2004emmy,trautman1967noe,modinoemmy}. This theorem introduces the condition of invariance to the form taken by a physical law with respect to any rotational or translational (spatial or temporal) transformation \cite{viazminsky2018framing}.
These invariances provide a method for calculating both symmetries and constants of motion with an arbitrary number of degrees of freedom \cite{trautman1967noe,modinoemmy}.
The different behaviors of the moving system can be contemplated as given by the geometrical nature of the system. The invariance condition is given by the conservation of constants in the symmetry of this system.

\section{Methods}

A study of the kinematic and dynamic behavior of the oscillator studied in a previous work \cite{ElenayJimena}, whose pendulum now is coupled to a moving platform with a certain slope alpha, is carried out. Due to this modification of the characteristics of the oscillator, its pendular motion is not perpendicular to the ground.

The first step is to define the mathematical formalism and the reference system. Subsequently, studying the mobile system as a flywheel, the value of alpha for which the system performs the maximum displacement will be found. On this basis, the different ways in which the introduction of a load can operate will be analyzed \cite{ElenayJimena}. Once these concepts are clarified, the general expression of a transformation matrix is arrived at.

\section{Results and discussions}

\subsection{Definition of reference system}

\begin{figure}[htb!]
\centering
\includegraphics[width=80mm]{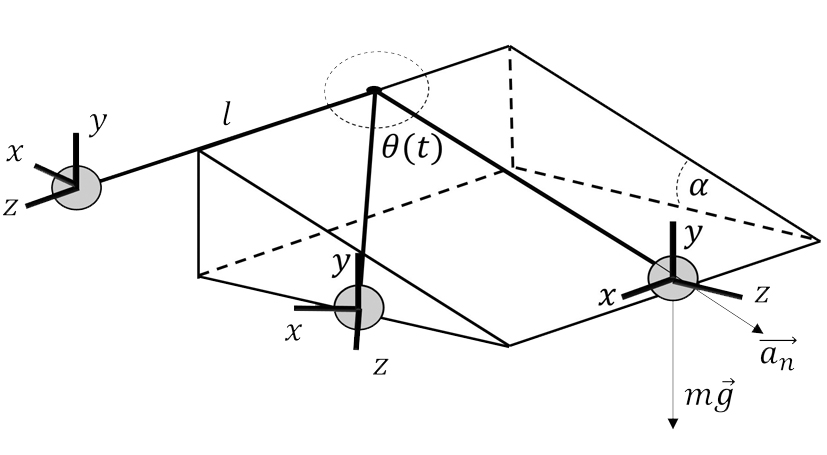}
\caption{Reference framework. Pendulum of mass $m$ with a normal acceleration ($\vec{a}_n$) rotating about $\theta=360^\circ$ and with an inclination $\alpha$.}
\label{Fig1}
\end{figure}

This system consists of a pendulum on a static plane with a certain slope given by the angle $\alpha$ (Fig.\ref{Fig1}), which has a mass $m$ and a rigid rod of length $l$ fixed to the mobile that oscillates on it with an angle $\theta$. The angle at which the pendulum starts its trajectory at $\theta_0$. The coordinate system is placed in the mass of the pendulum itself. The oscillator is defined in such a way that in the present analysis the friction with the air and the pendulum axis is depreciated. As a consequence, the oscillator will have a uniform angular velocity which will only be varied by the introduction of forces. With the axes defined, the forces due to normal acceleration and gravity are projected (Eq.\ref{eq:1}).

\begin{equation}
\left.\begin{matrix}

\vspace{5mm}
\vec{F_{x}}=(mg \sin\theta \sin\alpha)\;\widehat{x} \\ 
\vspace{5mm}
\vec{F_{y}}=(ml\omega^{2}\sin\alpha \cos \theta +mg)\; \widehat{y}\\
\vec{F_{z}}=((ml\omega^{2} +mg\cos\theta) \cos\alpha)\; \widehat{z}\\ 

\end{matrix}\right\}
\label{eq:1}
\end{equation}

%
%

\subsection{Mobile system displacement}


\subsubsection{Inclination that maximizes the displacement of the mobile}

On the one hand, a static system can be described (Fig.\ref{Fig1}), in which the pendulum motion of the system maintains a constant period due to the lack of energy losses because it is an ideal system. Therefore the oscillation of the pendulum will start at $\theta_0$ and go to $-\theta_0$, thus maintaining its motion at the angle $\theta$ without losing degrees in the oscillation.

The value of the inclination $\alpha$ for which the maximum value of kinetic energy is obtained is to be found, since a certain part of this energy can be released by the motion of the oscillator.

\begin{figure}[htb!]
\centering
\includegraphics[width=80mm]{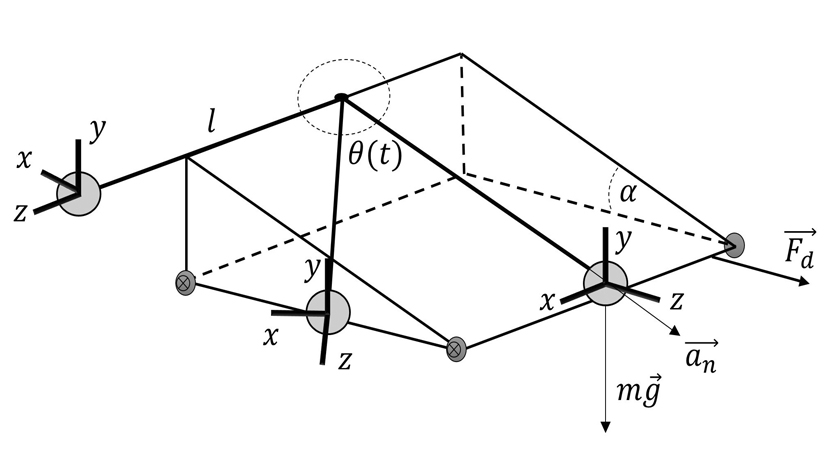}
\caption{Description of the reference framework with displacement. Pendulum of mass $m$ with a normal acceleration ($\vec{a}_n$) rotating about $\theta=360^\circ$ and with an inclination $\alpha$. $\vec{F_d}$ indicates the direction of displacement of the mobile.}
\label{Fig2}
\end{figure}

For this purpose, it is possible to make this system mobile by introducing fixed wheels oriented in the direction of the $z$-axis (Fig.\ref{Fig2}), where the maximum displacement of the mobile will allow to identify the optimal value of $\alpha$.

The mobile releases all its energy when the pendulum is at angle $\theta=0^\circ$, because for this angle its potential energy is minimum and the kinetic energy is maximum. It is possible to study the displacement given by the three axes of the force projections (Eq.\ref{eq:2}). 

\begin{equation}
\vec{F_{d}}=(\vec{F_{x}}\sin\theta +\vec{F_{z}}\cos\theta)\;\widehat{z}
\label{eq:2}
\end{equation}

The angular velocity at this point is obtained (Eq.\ref{eq:3}), using for its calculation the conservation of mechanical energy at the points of maximum and minimum potential. 

\begin{equation}
\omega=\sqrt{\frac{2g}{l}\sin\alpha (1-\cos{\theta_{0}})}
\label{eq:3}
\end{equation}

To calculate the force that will maximize the displacement, the angular velocity where the potential energy is minimum, with an initial angle $\theta_0=90^\circ$ (Eq.\ref{eq:4}).
It will be the derivative as a function of $\alpha$ of Eq.\ref{eq:4} that yields the result that maximizes this function for a given $\alpha$ (Eq.\ref{eq:5}). This yields the value of the slope that maximizes the displacement, $\alpha=36,37^\circ$.

\begin{equation}
\vec{F}_{d}=((2mg\sin\alpha +mg)\cos\alpha)\;\widehat{z}
\label{eq:4}
\end{equation}

\begin{equation}
 \frac{d\vec{{F}_{d}} }{d \alpha}=2mg\cos^2\alpha-2mg\sin^2\alpha-mg\sin\alpha=0
 \label{eq:5}
\end{equation}

\begin{figure}[htb!]
\centering
\includegraphics[width=80mm]{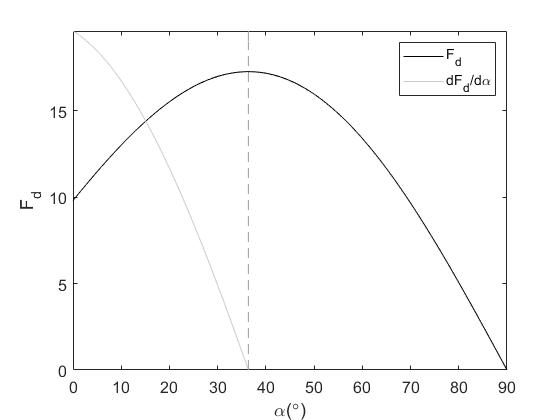}
\caption{Ratio of the displacement force ($\vec{F}_{d}$) and its derivative ($d \vec{F}_{d} /d \alpha$) at $\theta=0^\circ$ for the possible values of $\alpha$ considering $\theta_0=90^\circ$. For $\alpha=36.37^\circ$, $\vec{F}_{d}$ maximum is obtained.}
\label{Fig3}
\end{figure}

Given all possible values of $\alpha$, it is observed in Figure \ref{Fig3} that the maximum displacement is satisfied for $\alpha=36.37^\circ$.

\subsubsection{Inclination that maximizes the displacement of the mobile with a frictional force}

For the same conditions, the displacement force is analyzed there is a frictional force due to the ground (Ec.\ref{eq:6}).

\begin{equation}
\vec{F}_{d2}=(\vec{F_{x}}\sin\theta +\vec{F_{z}}\cos\theta -\vec{F}_{roz})\;\widehat{z}
\label{eq:6}
\end{equation}

Therefore, it is considered that there is a coefficient of friction $\mu$ of the mobile with a ground, and the instant of the movement in which the force of the pendulum in the direction of displacement is maximum is analyzed. ($\theta=0$) (Ec.\ref{eq:7}).

\begin{equation}
\vec{F}_{d2}=((2mg\sin\alpha +mg)\cos\alpha-\mu(m+M)g)\;\widehat{z}
\label{eq:7}
\end{equation}

\begin{equation}
 \frac{d\vec{F}_{d2}}{d \alpha}=2mg\cos^2\alpha-2mg\sin^2\alpha-mg\sin\alpha=0
 \label{eq:8}
\end{equation}

\begin{figure}[htb!]
\centering
\includegraphics[width=80mm]{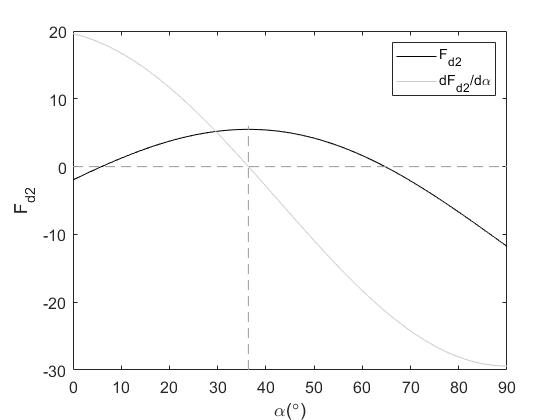}
\caption{Ratio of the displacement force with friction ($\vec{F}_{d2}$) and its derivative ($d\vec{F}_{d} /d \alpha$) at $\theta=0^\circ$ for the possible values of $\alpha$ considering $\theta_0=90^\circ$. For $\alpha=36.37^\circ$, $\vec{F}_{d2}$ maximum is obtained.}
\label{Fig4}
\end{figure}

As it can be seen, Eq.\ref{eq:8} renders the same result as Eq.\ref{eq:5}, determining the same value of $\alpha$ that makes the displacement of the mobile maximum. If this new displacement force is represented with respect to $\alpha$ (Fig.\ref{Fig4}), considering the friction it is observed that the curve has undergone a vertical translation that depends on $\mu$, $m$ and $M$, but the value of $\alpha$ that maximizes the displacement is the same as the one obtained in the calculation under ideal conditions $\alpha=36.37^\circ$.

\subsubsection{Displacement of the mobile as a flywheel}

The motion of the mobile is described taking as reference system the pendulum when it is at its maximum kinetic and minimum potential energy. The motion of the pendulum starts at an angle $\theta_0=90^\circ$ and moves in a negative direction on the $y$-axis, that is, in the positive direction of gravity, therefore, the mobile moves in the positive $z$ axis only when the pendulum is at the maximum kinetic energy ($\theta=0^\circ$). After that point, the pendulum continues its travel in the positive direction on the $y$-axis to its final position in the negative direction on the $z$-axis, which is that of the initial inverse angle ($-\theta_0$) minus the height lost by the displacement of the mobile. 
The energy is released only when the pendulum is at its maximum kinetic energy, so the mobile remains static and does not return to its initial position on the $z$-axis.

In case the pendulum was not part of a mobile system, but of a static system, that is to say in case it not have the wheels for its displacement, or in case the mass of the pendulum was of enough weight to prevent the system from moving, the energy would not be released in the form of displacement but it would remain in the system causing the oscillation to return to its initial inverse angle ($-\theta_0$) without having loss of height. It would thus preserve its initial equilibrium condition while keeping its period constant.

\begin{figure}[htb]
\centering
\includegraphics[width=90mm]{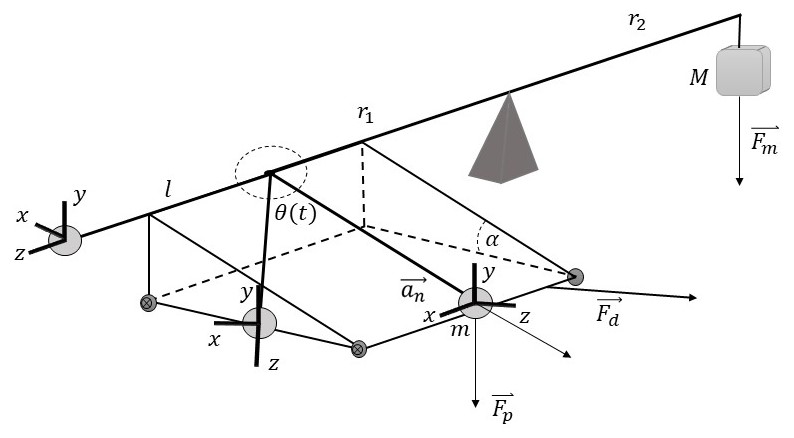}
\caption{Full oscillator description. System formed by a simple pendulum connected to a seesaw with arms $r_1$ and $r_2$. At the end of arm $r_1$ there is a pendulum of mass $m$ with a normal acceleration ($\vec{a}_n$) rotating about $\theta=360^\circ$ and with an inclination $\alpha$. $\vec{F_d}$ indicates the direction of displacement of the mobile. At the end of the arm $r_2$ there is the variable mass $M$ with weight $\vec{F_m}$.}
\label{Fig5}
\end{figure}

Knowing the angle $\alpha$ that maximizes the displacement of the mobile, now it can be taken into account that the pendulum system with mass $m$ is connected to a seesaw in other end there is a variable mass $M$ with only vertical motion (Fig.\ref{Fig5}). The system of mass $m$ is considered as a flywheel, because the kinetic energy is released at the point of minimum potential energy, i.e. for $\theta=0^\circ$, where the linear momentum of the system has to be conserved. 

In order for the system to be in equilibrium, the torque experienced by both ends at $\theta=0^\circ$ must be compensated (Eq.\ref{eq:9}).

\begin{equation}
r_{1}\left(\frac{1}{2}ml\omega^2+mg\right)=r_{2}Mg\\
\label{eq:9}
\end{equation}


For $l=r_1=r_2=\unit[1]{m}$, we have an equivalent mass $M$ of $1.5m$. This transfers an energy while conserving the momentum, which implies a decrease in the angular velocity of the pendulum and an increase in the linear velocity of the mobile. Given these conditions, the linear velocity of the system is obtained when it is at the point of minimum potential, defined by the angular values $\theta=0^\circ$ and $\alpha=36.37^\circ$ (Eq.\ref{eq:10}).

\begin{equation}
\left.\begin{matrix}
1.5m\cdot v_{p}=(m+M)\cdot v_{m+M}\\
v_{m+M}=\frac{1.5m(\omega l)}{m+M}
\end{matrix}\right\}
\label{eq:10}
\end{equation}
\hspace{2mm}

The function of the displacement of the system due to the variation of the mass $M$ given by the seesaw system can be known only when the pendulum is at $\theta=0^\circ$ and with a slope $\alpha$. Due to the inclination of the moving system, the direction of the pendulum's inertia is not the same as the direction of gravity.




\subsubsection{Implications due to mobile displacement}

The different apparent weight geometries of the system given by the different system variables describe the displacement of the system.

Initially a linear displacement loss will be reflected in a loss of angular velocity and potential energy. Therefore, this will result in a decrease in the height of the pendulum's oscillation. This decrease in energy is produced by the work done by the moving system due to the displacement (Eq.\ref{eq:11}).

\begin{equation}
W=F\cdot S=(F_{x},F_{y},F_{z})\cdot (0,0,s_z)
\label{eq:11}
\end{equation}

A decrease of energy in the form of height implies a variation of angular velocity. Due to the conservation of the mechanical energy of the system, it is possible to calculate the angular velocity by equating the maximum kinetic and potential energies (Eq.\ref{eq:12}), where $h_{2}$ is the new height given by the loss of potential energy. 

\begin{equation}
\omega _{2}=\frac{\sqrt{2gh_{2}}}{l}
\label{eq:12}
\end{equation}

As the system combines a pendulum with a seesaw (Fig.\ref{Fig5}), it is possible to transform this velocity increase into a force applied to the system: concretely, velocity variation implies a variation of apparent weight, and this implies a higher normal force on the normal force of the system.

This result would correspond to the angular velocity that would have to be added to the initial one to counteract the loss of angle or height due to the displacement of the mobile. The way to add this new angular velocity is by means of a second force (Eq.\ref{eq:13}) applied tangentially to the motion of the mobile and in a normal directo with regard to to the oscillation of the pendulum.

\begin{equation}
{F_{i}}_{2}=ml\omega_2^2
\label{eq:13}
\end{equation}

So to counteract this angular variation a force is exerted on the $y$-axis. That is, the initial one (Eq.\ref{eq:1}) but taking the initial values of $\theta_{0}=-180^\circ$ $\theta=0^\circ$ and $\alpha =30^\circ$. Applying Ec.\ref{eq:13}, we obtain Ec.\ref{eq:14}.

\begin{equation}
{F_{i}}_{2}=ml\omega ^{2}\sin \alpha \cos \theta +mg+ml\omega_2^2
\label{eq:14}
\end{equation}

It is possible to express this force variation as a function of the apparent weight (Eq.\ref{eq:15}) by equating the torque offset (Fig.\ref{Fig5}) and taking Eq.\ref{eq:14} as the mass weight $m$. Thus leaving the expression as a function of the angular positions $\alpha$ and $\theta$.

\begin{equation}
\frac{M}{m}=2\sin^2\alpha\cos\theta+1+2h_2
\label{eq:15}
\end{equation}

In this way, the loss of energy, in the form of displacement of the mobile in the z-axis, is related to a variation of force applied tangentially to the motion of the mobile and normal in direction with regard to the oscillation of the pendulum. In turn, this variation of force translates into an increase of the apparent weight necessary to maintain the initial symmetry of the system.





%
%

\subsection{Apparent weight variation versus system geometry}

Once the angle $\alpha$ that maximizes the displacement of the mobile is known, it is possible deepen the analysis of the working of the assumed system: a pendulum in equilibrium relation with a seesaw so that the mass $M$ must adapt to the variations of apparent weight of the pendulum along the oscillation $\theta$ (Fig.\ref{Fig5}). The torque at both ends is compensated (Eq.\ref{eq:16}), so that the expression for apparent weight variation in the system (Eq.\ref{eq:17}) is obtained for $\theta_0=180^\circ$ and $r_1=r_2=\unit[1]{m}$.

\begin{equation}
r_{1}\cdot (ml\omega ^{2}\sin \alpha \cos \theta +mg)=r_{2}\cdot (Mg) \\
\label{eq:16}
\end{equation}
\begin{equation}
\frac{M}{m}=4\cos \theta \sin ^{2}\alpha +1
\label{eq:17}
\end{equation}
\hspace{2mm}

Therefore, the apparent weight variation depends on the geometrical position in which the pendulum is located, which is given by the angular parameters $\alpha$ and $\theta$. 

This apparent weight variation is usually applied to the weight of the masses, since it is the masses that vary as angular parameters are modified. It may involve mass variations large enough to overcome the tangential forces due to inertia and thus keep the system static while maintaining symmetry in the system. 

Conversely, these variations are also related to energy losses, since if the tangential or inertial forces exceed those of apparent weight, the mobile moves, exerting a work on the system.

\subsection{System transformation matrix}


The system under study can be approached from a theoretical standpoint, understanding the Galileo and Lorentz transformations from the Noether theorem and using, therefore, a proper system of $M/m$ coordinates, $\theta _{0}$, $\alpha$, $\theta$ (Eq.\ref{eq:17}). In this way an analogy with the Lorentz system is achieved, the coordinates being invariant $\theta _{0}$ and $\alpha$, and the variables $\theta$ and $M/m$.  
In this way, the transformation matrix is constructed based on these coordinates, which will allow to determine in which point the system will be found when varying any coordinate.

Each system has a lagrangian that allows to obtain the equations of its motion. From this lagrangian it is possible to determine the time evolution, the conservation laws and other important properties of the system. For example, in order to know the existence of symmetries in the system, one must study the conserved magnitudes.

It can be stated that there is a strong relationship between the continuous symmetries given by a system and the existence of its conservation laws. Emmy Noether formalized this theory in a theorem, which explains that if the physical laws do not change under spatial translations, linear momentum is conserved; and, if they do not change under rotations, both angular momentum and energy are conserved \cite{lemos2007mecanica}. That is, if the lagrangian of the system is invariant, not changing under continuous symmetry, then there is an associated conservation law \cite{zee2013einstein}. If we have a time derivative equal to zero, i.e., a magnitude that does not change in time, for each continuous $\delta_x$ symmetry we have a conserved magnitude $\frac{\partial \mathcal{L}}{\partial v} \delta_x = Q = \text{constant}$.

For the case where the canonical transformations are time-independent, they take a new notation, of the form: $\eta = F(\varepsilon)$. 
Let $M$ be the transformation Jacobian matrix, with elements $M_{ij}=\frac{\partial \eta _{i}}{\partial \varepsilon _{j}}$, the determinant of $M$ must be equal to 1. The condition for the transformation to be canonical is that the fundamental matrix $J$. Since $J$ is the transformation matrix also called generator, it must fulfill that it is antisymmetric $J=-J^T$.

It is a necessary and sufficient condition that $M^T JM=J$. In such case the system's symmetry would be conserved under rotations. The possible variations introduced modify parameters in such a way that the system tries to compensate them to preserve its symmetry.

The idea of finding a transformation matrix as a way of dealing with the passage from one reference system to another, in our case from non-inertial to inertial, has been developed since the $16^{th}$ century. The main precursor of this idea was Galileo \cite{galileo}, but he was not the only one to deal with it. Centuries later, Lorentz followed in his footsteps \cite{Lorentz}, completing and generalizing his transformations. He found a mechanism of transformation of values between reference systems with relative motion with velocity $u$ but with a maximum velocity $c$ equal for those reference systems \cite{tiberius}.

Lorentz transformations relate the measurements of a physical quantity made by two different inertial observers, and represent the point in space-time (determined by an instant of time and three spatial coordinates) thus preserving the value of the speed of light constant for all observers.

It is for this reason that Noether established, while investigating questions of special relativity, that to each infinitesimal transformation of the Lorentz group corresponds a conservation theorem \cite{noether2004emmy}.

Therefore, by analyzing Eq.\ref{eq:17} for different angles, different linear dependencies are obtained (Eq.\ref{eq:18},Eq.\ref{eq:19}). In particular for a complete oscillation of 360$^\circ$, and fixing the vertical tilt plane at $\alpha=90^\circ$, the minimum potential with $\theta_{0}=-180^\circ$ is taken as the point of analysis

\begin{equation}
\frac{M}{m}\sim \frac{77}{25}\alpha + \frac{29}{50}\quad\text{if $\theta=0$}
\label{eq:18}
\end{equation}

\begin{equation}
\frac{M}{m}\sim \frac{77}{25}\theta - \frac{27}{2}\quad\text{if $\alpha =90$}
\label{eq:19}
\end{equation}

These lines correspond to the linear regressions resulting from representing $M/m$ versus the different angles $\theta$ and $\alpha$. This shows that, depending on the quadrant in which it is, the slope could vary its sign. This is related to how the apparent weight is acting, whether its function is to accelerate or slow down the system. The relationship between the pendulum and the seesaw reveals that the apparent weight is greatest when the system is at minimum potential energy. This occurs because of the relationship between the inertia and apparent weight of the system. 

Representing these dependencies, it can be see that they correspond to different planes: on the one hand there is a plane showing the variation of $M/m$ with respect to $\theta$ and on the other the variation of $M/m$ with respect to $\alpha$. The union of these planes gives rise to a three-dimensional image of the variation of the apparent weight, being able to observe at the same time, in a single graph, the correspondence of the value of $M/m$ with these two angular positions. 

Analogous to the Lorentz transformations, which relate the measurements of a physical quantity made by two different inertial observers and represent the point in space-time (determined by an instant of time and three spatial coordinates), thus keeping the value of the speed of light constant for all observers, in the transformation matrix, the space will be represented as a system of the angles $\theta$ and $\alpha$, preserving the magnitude of the apparent weight.

Starting from Eq.\ref{eq:17}, the representation of all possible angles can be expressed in matrix form. However, the limit of expression is marked by the boundary of the system, and therefore the representation will not be marked in the plane $\mathbb{R}$, but in Eq.\ref{eq:20} we resort to a representation with complex numbers given by the plane $\mathbb{C}$, thus obtaining a matrix of n rows by m columns. In this example, if $\alpha$ varies between 0º-180º and $\theta$ between 0º-360º  a matrix of dimension 181x361 is obtained. Thus for each value of $\alpha$ there are 361 values of $\theta$.

\begin{equation}
\frac{M}{m}\hspace{-0.05cm}=
\resizebox{0.83\hsize}{!}{%
$\displaystyle
\begin{pmatrix}
1-\frac{1}{2}(e^{i\theta}+e^{-i\theta})(e^{-i\alpha}-e^{i\alpha})^2 & 
\cdots &
1-\frac{1}{2}(e^{i(\theta+m)}+e^{-i(\theta+m)})(e^{-i\alpha}-e^{i\alpha})^2 \\
\vdots & 
\ddots & 
\vdots \\
1-\frac{1}{2}(e^{i\theta}+e^{-i\theta})(e^{-i(\alpha+n)}-e^{i(\alpha+n)})^2 &
\cdots  &
1-\frac{1}{2}(e^{i(\theta+m)}+e^{-i(\theta+m)})(e^{-i(\alpha+n)}-e^{i(\alpha+n)})^2
\end{pmatrix}
$}
\label{eq:20}
\end{equation}
\vspace{3mm}

\begin{figure}[htb!]
\centering
\includegraphics[width=100mm]{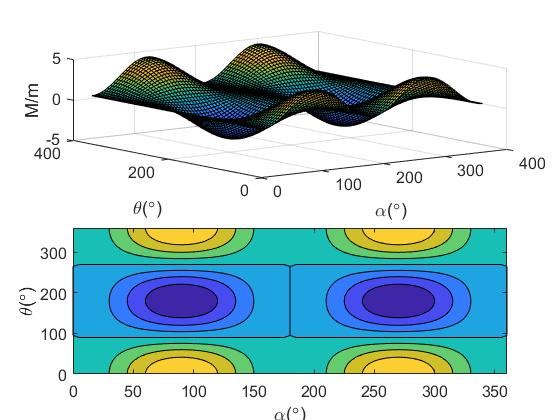}
\caption{Representation in $R^3$ of the variation of the apparent weight with respect to $\theta$ and $\alpha$. A two-period representation of the system is observed.}
\label{MATRIZ 1}
\end{figure}

The three-dimensional representation is nothing more than the graph of the matrix expression of the apparent weight (Fig.\ref{MATRIZ 1}). It is observed that depending on the position of the angles $\theta$ and $\alpha$ a value of the apparent weight will be obtained, being able to have positive maximum and negative minimum values, can be represented as peaks of energetic expression. A positive value represents the need to provide energy, and a negative value that of energy extraction, all this to preserve the system's symmetry. 
These values are periodic since they will be given every 180$^\circ$ for each angle due to the limits marked by the boundary of the system. However, they can be represented for infinite angles, since the system can chain periods indefinitely.



%
%

\subsection{Interpretation of a transformation matrix}

The Fig.\ref{MATRIZ 1} is represented by the conserved variables of $\alpha$ and $\theta$ by which the apparent weight of the system is represented. That is, due to the degrees of freedom of the system, two conserved magnitudes have been found that make its symmetry to be maintained. Therefore, in the case of altering the nature of the system by modifying some parameter, the system will compensate its apparent weight in order to preserve the symmetry.

The introduction of energy in the form of velocity can be an option to clearly visualize the different periods, since these have slight variations depending on whether an increase or a decrease of this velocity is obtained in the system. 
If the angular velocity is increased, the system is accelerating, so the period will decrease since it is inversely proportional to the angular velocity. 

Newton introduced in his book \textit{De methodis serierum et fluxionum} the concept of fluxion of the source, i.e. acceleration, which can also be expressed as the derivative of the velocity with respect to time \cite{newton1967methodis}. It is here that the concept of derivative appears for the first time, which makes us consider Newton as one of the fathers of calculus. It can be extracted from this concept that the more space we have, the more velocity can be obtained, thus reiterating that velocity is not speed but the distance given in a given time, so in this system the time can be interpreted as the point given by the two angles $\theta$ and $\alpha$. In other words, knowing the angles, the position of the pendulum and the apparent weight it carries an be known too.




This geometric object has a fractal form, whose structure is basic, fragmented and can be repeated at different scales due to an introduction of energy such as an angular variation. On the contrary, if the system did not have any energy variation, it would have a scale invariance, that is to say, due to the conserved variables of the system it would be an invariant and periodic system.



Knowing the conserved variables described by Noether's theorem, these can be interpreted not only as a spatial part in the system, but also as a temporal part. These conserved variables, even if bounded by their limit angles, can be extrapolated to infinite angles given by the number of repetitions of the period in the system. This means that, although its area is limited, it has a path or perimeter which can be infinite, and therefore it needs an infinite time to be reached. So we can liken this interpretation to that of a fractal figure, whose area is finite but its perimeter or path is given by an infinitesimal space \cite{baliarda2000koch}.

Collecting the whole expression in a transformation matrix makes both the visualization and the clarification of the concepts of energy input or extraction more feasible.

The analogies with the seemingly paradoxical relationship between acceleration, space and time, or with the seemingly paradoxical properties of fractal figures, that have been proposed can help to understand both the meaning of the matrix and its possible extrapolations, as long as all the above mentioned requirements are met (i.e. the Lagrangian is invariant to time transformations, that the transformation matrix is antisymmetric, and that the condition $M^T JM=J$ is fulfilled).

\section{Conclusions}

A variation of parameters such as displacement, friction or variation in height can mean a loss of energy in the system. 
Noether's theorem reflects the expression of the appearance of conserved magnitudes that do not depend on time, implying that the behavior of the system before any variation is defined by the natural tendency to conserve its symmetry. 
The introduction or variation of parameters in the system will cause a modification or compensation in its behavior in such a way that as it will tend to remain in its equilibrium state it will seek a compensation that makes it maintain its symmetry. 
So knowing the values of $\alpha$ and $\theta$, and knowing that they are the conserved variables of the system, it would be possible not only to have the information about the apparent weight variable, but also to know both the space in which they are located, as well as their temporal representation. The limit of the geometric expression makes the representation of these variables to be bounded. This limit can be taken to infinity due to the periodic repetition given by the geometry of the system. 
This representation means that although the system has a finite surface, its path can become infinite and therefore, it needs an infinite time to be reached. A possible analogy could be that of a fractal figure, where even having a finite area we can have an infinite path that can be given by an infinitesimal space. 
We can clarify all these concepts in a transformation matrix, which also helps to extrapolate them to other systems that satisfy the same conditions. Therefore, when faced with an asymmetric system, we look for the conserved variables that comply with Noether's theorem to know the existence of symmetries in the system, encompassing both rotations and transformations, therefore of being able to obtain a clear visualization of the values expressed and thus the need to provide or extract energy to the system.


\bibliography{mybibfile}

\end{document}